\documentclass{aa}

\usepackage[varg]{txfonts}
\usepackage{natbib}
\usepackage{graphicx}
\usepackage{times}
\usepackage{color}
\usepackage{hyperref}
\hypersetup{
  hidelinks,
  colorlinks=true,
  linkcolor=blue,
  filecolor=blue,
  urlcolor=blue,
  citecolor=blue,
}
\usepackage{bm}
\usepackage{enumitem}   
\usepackage{scalerel}
\usepackage{multirow}
\usepackage{stfloats}
\usepackage{float}

\usepackage{tabularx}
    \newcolumntype{C}{>{\centering\arraybackslash}X}
    \newcolumntype{L}{>{\raggedright\arraybackslash}X}
    \newcolumntype{R}{>{\raggedleft\arraybackslash}X}
\usepackage{array}

\usepackage{siunitx,booktabs}
\usepackage[version=4]{mhchem}
\usepackage{collcell}

\bibpunct{(}{)}{;}{a}{}{,}
\begin{document}

   \title{Unveiling the spectacular over 24-hour flare of star CD-36 3202}

   \author{K. Bicz\inst{1, 2} \and R. Falewicz\inst{1, 2} \and M. Pietras\inst{1}}

   \institute{Astronomical Institute, University of Wroc\l{}aw, Kopernika 11, 51-622 Wroc\l{}aw, Poland
              \and
             University of Wroc\l{}aw, Centre of Scientific Excellence - Solar and Stellar Activity, Kopernika 11, 51-622 Wroc\l{}aw, Poland\\
             \email{bicz@astro.uni.wroc.pl}
             }
 
  \abstract{We studied the light curve of the star CD-36 3202, observed by TESS for the presence of stellar spots and to analyze the rotationally modulated flare. We mainly wanted to model the light curve of this flare and estimate its location regarding stellar spots. The flare lasted approximately 27$\,$h. Using our tool new \texttt{findinc\_mc} we managed to estimate the inclination angle of the star to $70^\circ\pm8^\circ$. With \texttt{BASSMAN} we modeled the light curve of the CD-36 3202 and we estimated that three spots are present on the surface of this star. The mean temperature of the spots was about $4000\pm 765\,$K, and the total spottedness was on average $11.61\%\pm0.13\,$\%. We created a new tool named \texttt{MFUEA} to model rotationally modulated flares. Using this software we estimated the latitude of the flare long-duration event equal to $69^{+2}_{-1}\,$deg in latitude. Our estimation of the flare's location was the first recreation of the exact position of a flare compared with the spots. The flare is placed 12$^\circ$ from the center of the coolest spot. This makes the flare related to the magnetic processes above the active region represented by the spot. Removing the effects of rotational modulation from the flare light curve allowed us to correct the estimation of bolometric energy released during the event from $(1.15\pm 0.35)\times 10^{35}\,$erg to $(3.99\pm 1.22)\times 10^{35}\,$erg.}


   \maketitle
\vspace{-1.1cm}
\section{Introduction}
In the vast expanse of the universe, stars exhibit a fascinating array of behaviors and phenomena. Among these captivating features are starspots, stellar flares, and stellar activity, which provide insights into the dynamic nature of stars \citep{Howard_2019, Roettenbacher_2018}. It is an umbrella term used to describe the range of activities resulting from the interaction between a star's magnetic field and its internal dynamics. Stellar activity is highly variable, with different stars exhibiting varying degrees of magnetic activity \citep{Yang_2017, Mathur_2014}. Younger stars, for instance, tend to be more active, displaying frequent flares and larger starspots, while older stars generally exhibit less activity \citep{Davenport_2019}.

Starspots, akin to sunspots on our own Sun, are dark, cooler regions that appear on the surface of stars \citep{Biermann_1941,Hoyle_1949,Chitre_1963,Bray_and_Loughhead_1964,Deinzer_1965,Dicke_1970}. These spots are caused by intense magnetic activity and can vary in size, shape, and duration. They are commonly found on stars that possess a convective envelope, where hot plasma rises from the interior, next cools, and finally descends back into the star. The presence of starspots indicates the presence of strong magnetic fields on these stars, affecting their overall brightness and spectral characteristics \citep{Flores_2022}.

Stellar flares are sudden and dramatic releases of magnetic energy in the form of intense bursts of electromagnetic radiation. Stellar flares are analogous to solar flares but can be significantly more powerful \citep{Namekata_2017,Pietras_2022}. They occur when the magnetic field lines of a star become twisted and tangled, leading to a rapid reconfiguration that releases tremendous amounts of energy. Stellar flares emit radiation across the electromagnetic spectrum, including X-rays and ultraviolet light, and can be detected by specialized observatories. These energetic events provide a glimpse into the violent and dynamic processes taking place within stars.

Understanding starspots, stellar flares, and stellar activity is crucial for comprehending the physical properties and evolutionary processes of stars. Through the study of these phenomena, we can gain insights into stellar magnetism, stellar winds, and the impact of stellar activity on planetary systems. Observing starspots and stellar flares on distant stars, enables us to compare them to our own Sun's activity, and allows us to explore the broader context of stellar behavior and the potential habitability of exoplanets \citep{Gunther_2020,Bogner_2022}.

Here, we present our analysis of the starspots distribution on CD-36 3202 and the analysis of the long-duration event (LDE) that occurred on that star. We use light curves from TESS and we compare the results with the earlier published papers. We estimated the number of spots on this star and their parameters (temperature, size, stellar longitude, and latitude). We estimated the location of the flare and how it is arranged compared with the locations of the spots.
In Section \ref{sec:cd} we describe the star itself and in Section \ref{sec:data} we describe the observations. Our starspots modeling and flare analysis methods are in Section \ref{sec:spots} and Section \ref{sec:flares} respectively. The results are presented in Section \ref{sec:results} and a discussion and conclusions are provided in Section \ref{sec:conclusion}.

\section{CD-36 3202}\label{sec:cd}
We analyzed the variability and very interesting flare on this star. 
CD-36 3202 (also known as TIC156758257) is a partially convective K2V \citep{Torres_2006} young star (age of around 40 Myr \citep{bell_2014}) at the distance $90\,$pc, with the mass $0.8\,$M$_\odot$. Its radius is $0.8\,$R$_\odot$, the effective temperature 4885$\,$K (MAST catalog\footnote{http://archive.stsci.edu}), and the estimated rotational period equal 0.23536 $\pm$ 0.00046 days. For the analysis of variability of this star we estimated its inclination angle $i = 70^\circ \pm 8^\circ$ (for more details see Appendix \ref{appendix:inc}).

\begin{figure*}[!b]
    \centering
    \includegraphics[width=\textwidth]{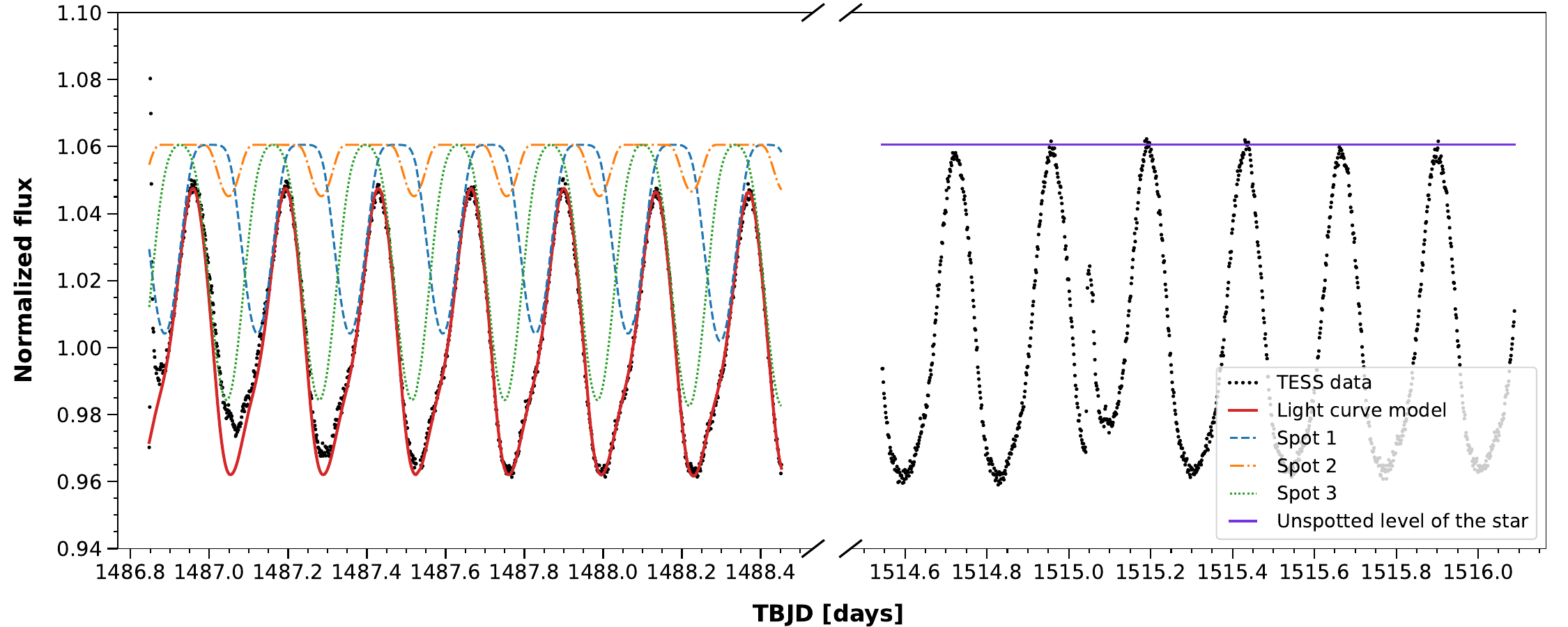}
    \caption{TESS light curves from sectors 6 and 7, with a reconstructed starspot model and the unspotted stellar level highlighted. Left side: the part of the TESS light curve from sector 6 (black dots) during which the long-duration flare occurred, with the starspot model that recreates the curve (red line). The green, blue, and orange curves present the contribution of individual spots to the whole light curve. Right side: the part of the TESS light curve from sector 7 (black dots) with the estimated unspotted level of the star marked with a purple line.}
    \label{fig:modeliampl}
\end{figure*}

The level of the flux of the unspotted CD-36 3202 is a crucial factor to conduct a proper analysis of the distribution of starspots. We used the TESS observations from sector 7 due to the occurrence of the highest values of the star's signal without any flares and the methodology presented by \citet{Bicz_2022} to evaluate this level and we received the value 1.06053 in normalized flux (see the right side of Figure \ref{fig:modeliampl}). 

Due to the flaring activity of the star, fast rotation period, and being a very young object we try to explain the variability by the presence of high spot activity on the star's surface (the percentage change in the luminosity is about 10\%).

Given the star's flaring activity, rapid rotation period, and youthful age, we are attempting to explain the observed variability through the heightened spot activity on the star's surface. This phenomenon is reflected in a notable alteration in luminosity, amounting to approximately 10\%.

\section{White light data}\label{sec:data}
In our investigation, we utilized data gathered from the TESS (\textit{The Transiting Exoplanet Survey Satellite}) \citep{Ricker_2014} mission. TESS, a space-based telescope, was launched in April 2018 and positioned in a highly elliptical orbit that completes the full cycle every 13.7$\,$days. The primary objective of the mission is to provide uninterrupted observations of a significant portion of the celestial sphere that is divided into 26 sectors. The satellite diligently monitors designated stars, employing a two-minute cadence (referred to as short cadence) and a twenty-second cadence (known as fast cadence) throughout the approximately 27-day monitoring period in each sector. Additionally, by utilizing Full Frame Images (FFI), one can obtain light curves with a cadence of 30 minutes. In order to maintain consistency in our light curve analysis, we specifically utilized the two-minute cadence light curves with the quality flag set to 0, ensuring the utilization of the most accurate and reliable data.

\section{Starspots modeling}\label{sec:spots}
We used BASSMAN\footnote{https://github.com/KBicz/BASSMAN} (Best rAndom StarSpots Model cAlculatioN) presented by \citet{Bicz_2022} to model starspots. 
\texttt{BASSMAN} recreates the light curve of a spotted star by fitting the spot(s) model, described by amplitudes, sizes, stellar longitudes, and latitudes of spot(s) to the data. 
To do so \texttt{BASSMAN} uses Broyden–Fletcher–Goldfarb–Shanno (BFGS) algorithm \citep{Flet87} and Markov chain Monte Carlo (MCMC) methods. The program uses the following ready-made software packages to model the spots on the star: starry \citet{Luger_2019}, PyMC3 \citet{Salvatier_2016}, exoplanet \citet{exoplanet:exoplanet}, theano \citet{exoplanet:theano}. 
Each star is expressed in \texttt{BASSMAN} as a linear combination of spherical harmonics. The star is described by the vector of the spherical harmonic coefficients (indexed by increasing the degree $l$ and order $m$). Each starspot on the star is the spherical harmonic expansion of a Gaussian, with the assumption that the spot is spherical.

It is worth mentioning, that the ,,starspots" which \texttt{BASSMAN} recreates are not exactly sunspot-like structures. The software approximates more probably the whole active region which consists of several individual spots by one ,,spot". There is no way to distinguish between these cases from light curves observed only in a single bandpass.


\section{Analysis of the flares}\label{sec:flares}
We used an automated, three-step software WARPFINDER (Wroclaw AlgoRithm Prepared For detectINg anD analyzing stEllar flaRes) presented by \citet{Bicz_2022} and \citet{Pietras_2022} to analyze and detect flares on the light curve. The first step is the method based on consecutive de-trending of the light curve described by \citet{Davenport_2014}. The second method of finding stellar flares is the Difference Method, based on \citet{Shibayama_2013}. The times from the lists of flares candidates created by both methods are compared and only the common detection in a designated time is treated by the software as a potential detection and passed on to the next level of verification. Then, we fit the assumed flare profiles \citep{Bicz_2022, Pietras_2022} to the observational data and check its quality with $\chi^2$ statistic. We use the probability density function of F-distribution to distinguish a stellar flare from data noise. Additionally, we calculate the skewness of the flare profile and check the bisectors of the profile (in order) to reject false detections. We assume that a stellar flare should have a rise time shorter than a decay time. We also reject all events with a duration of less than 6 observational points (12 minutes) for the two-minute cadence data and less than 18 observational points (6 minutes) for the 20-second cadence data. \texttt{WARPFINDER} uses two methods to estimate the flare's energy. The first is based on the method presented by \citet{Kovari_2007} and the second is based on the method presented by \citet{Shibayama_2013}.

To estimate the possible location of the flare on the star and fit the theoretical flare profile proposed by \citet{Gryciuk_2017} we created a new tool written in Rust named \texttt{MFUEA}\footnote{https://github.com/KBicz/MFUEA}. \texttt{MFUEA} models the modulated flare using the Differential Evolution method (DE). DE belongs to the category of evolutionary algorithms designed to address optimization problems with real-valued parameters. It serves as a highly effective method for black-box optimization, particularly when dealing with complex functions where finding the appropriate derivatives required to compute extreme values is challenging. DE takes inspiration from biological evolutionary processes, namely mutation, crossover, and selection. To achieve the desired solution, three crucial parameters come into play: population size, mutation factor, and crossover factor. The key features of DE can be summarized as follows:
\begin{enumerate}
    \item Unlike genetic algorithms \citep{Feoktistov_2004}, DE employs arithmetic combinations of individuals for mutation rather than relying on small gene perturbations.
    \item DE operates with three distinct populations of individuals: parents, trial, and descendants, all of which are of equal size.
    \item The size of the populations (representing potential solutions) typically remains constant throughout the evolution process.
    \item DE utilizes floating-point numbers, simplifying the implementation of crossover and mutation.
\end{enumerate}

The code expresses the flaring area as the photospheric footpoint of the flaring magnetic loop, from which the white light emission originates. It models the flux using the spherical geometry in the below equation:

\begin{table*}[!b]
\caption[]{Mean model of starspots on CD-36 3202 for the sector 6.}\label{tab:spots_cd}
\centering
\small
\begin{tabular*}{\hsize}{@{\hspace{0.2cm}}@{\extracolsep{\fill}}ccccc}
\hline\hline\noalign{\smallskip}
Spot & Spot relative & Spot size & Mean spot temperature & Spot latitude\\
number & amplitude [\%] & [\% of area of star] & [K] & [deg]\\
\noalign{\smallskip}\hline\noalign{\smallskip}
        1 & $1.50\,\pm\,0.15$ & $3.83\,\pm\,0.2$ & $4038\,\pm\,669$ & $\,\,\,52\,\pm\,2$\\
        2 & $1.83\,\pm\,0.37$ & $3.85\,\pm\,0.2$ & $4057\,\pm\,735$ & $-55\,\pm\,3$ \\
        3 & $2.53\,\pm\,0.15$ & $3.93\,\pm\,0.3$ & $3906\,\pm\,852$ & $\,\,\,62\,\pm\,2$ \\
\noalign{\smallskip}\hline\noalign{\smallskip}
\end{tabular*}
\end{table*}

\begin{equation}
    F_{\mathrm{modulated}}(\phi,\theta,t) = F_{\mathrm{model}}\cdot\left(\sin(\phi)\cos(i)+\cos(\phi)\sin(i)\cos(\theta-t)\right),
\end{equation}
where $\theta$ is the stellar longitude, $\phi$ is the stellar latitude, $F_{\mathrm{model}}$ it the flare profile from \citet{Gryciuk_2017}, and $t$ is the time-like array.
The footpoint is expressed as a set of uniformly distributed points in the circular area defined by the radius of this footpoint expressed as below:
\begin{equation}
    \hfil\hfil r = \sin^{-1}\left(\sqrt{\frac{A\cdot L_*}{\pi R_*^2 F_f(T_f)}}\right),
\end{equation}
where $A$ is the amplitude of the flare, $L_*$ is the luminosity of the star, $R_*$ is the radius of the star, $F_f$ is the flux of the flare with the temperature of the flare $T_f$. This expression can be derived from the energy of the flare estimation method presented by \citet{Shibayama_2013}. The parameters that the software fits are the latitude and longitude of the flaring area, parameters of the fitted profile, and the inclination angle of the star if the user needs it. In the beginning, the population is initialized. In our case, each solution (individual) represents the flare profile and the location of the flaring region. The individuals consist of genes. This means that a few genes are related to the parameters of the profile of the flare and a number of genes corresponds to the location of the flaring area. The number of genes corresponding to the location depends on the number of points that covers the entire flaring area.
It is crucial to have the appropriate range of values for each parameter (gene), and the initial population should encompass the entire search space to the greatest extent possible. In our specific scenario, each solution consists of a randomly generated set of vectors. Then, the population undergoes a mutation process, wherein a trial vector is created for each individual in the parent population. Exactly one mutated individual must be generated for each parent individual.

To enhance the diversity of perturbed parameter vectors, the population of mutated vectors undergoes a crossover procedure. Through recombination, the crossover combines elements from the parent vector and the mutant vector to construct trial vectors.

To determine whether an individual should be included in the next generation, the trial vector is compared to its direct parent. Suppose an individual from the trial vector exhibits an equal or lower objective function value than its parent vector. In that case, it replaces the corresponding individual in the parent vector for the next generation. This evolutionary process is repeated until the convergence criterion is met.

\section{Results}\label{sec:results}

TESS observed CD-36 3202 in sector 6. In this period, the variability of this star was not constant in time so we divided the whole sector into 13 parts where the variability was approximately constant. The division allowed us to create consistent models that can represent the evolution of spots in the whole sector and eliminate solutions that could describe only one part of the sector or were unequivocal. We obtained the three-spots model with spots separated by $82^\circ\pm3^\circ$ and $47^\circ\pm4^\circ$ in the longitude from the middle spot or by $0.23\pm 0.01$ and $0.13\pm 0.01$ in a phase. The mean spot model for this sector can be seen in Table \ref{tab:spots_cd}. The contribution of each spot to the whole light curve and the model of the light curve can be seen on the left part of Figure \ref{fig:modeliampl}. The mean temperature of spots is $4000\pm765\,$K and the total spotedness of spots is $11.61\%\pm0.13\,$\% of the stellar disk. This fits quite well with the analytic solutions for the total spotedness ($14.04\pm0.76\,$\%) and the average temperature of the spots ($3622\pm 51\,$K) presented by \citet{Namekata_2017}. 

\begin{figure*}[ht!]
    \centering
    \includegraphics[width=\textwidth]{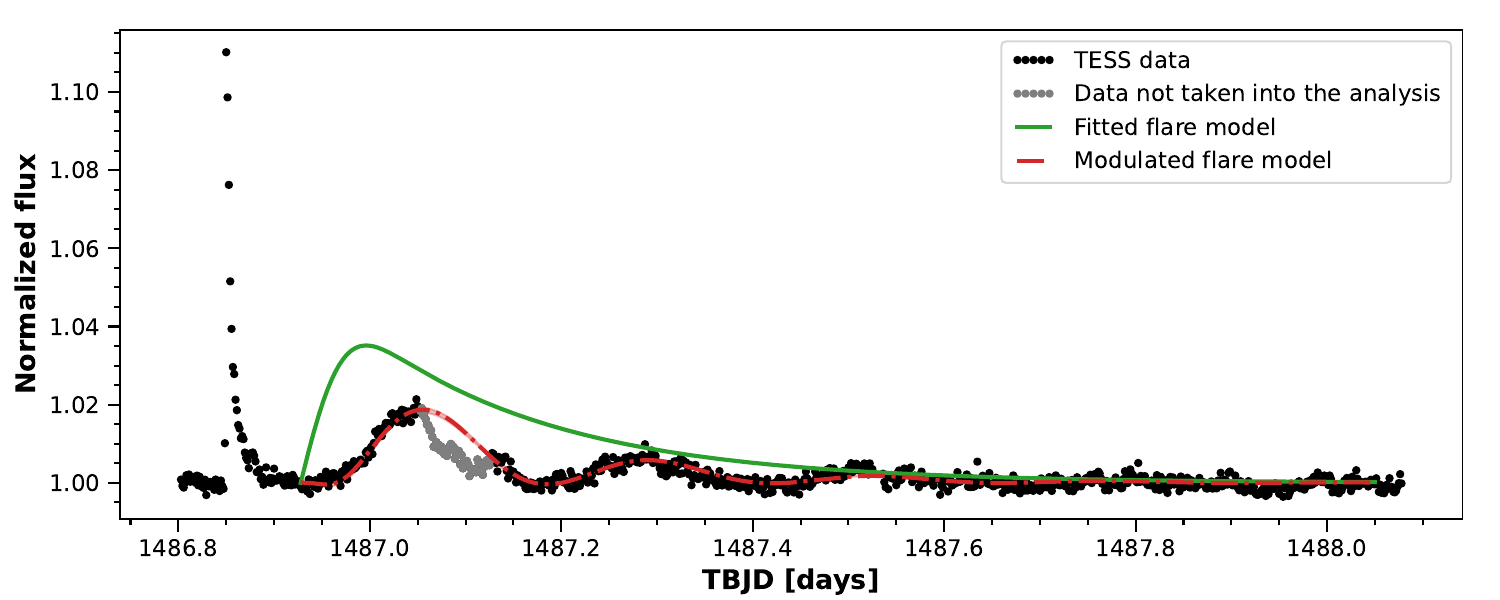}
    \caption{Part of the TESS light curve from sector 6 corrected to the variability caused by stellar
spots. The red curve and red interval mark the modulated flare model and the fit error respectively.
The green curve shows the recreated course of the stellar flare without the effects of foreshortening
and stellar rotation. The gray points mark the data not used in the analysis. The translucent red area presents the uncertainty area.}
    \label{fig:modulatedflare}
\end{figure*}

We managed to recreate the profile of the flare, its radius, and the astrographic position of the flare using \texttt{MFUEA}. During the calculations \texttt{MFUEA} included the quadratic limb darkening law with coefficients estimated by \citet{Claret_2017} for the temperature and the logarithm of the gravitational acceleration of this star. The flaring region was located on the $69^{+2}_{-1}\,$deg and had radius about 3.14$\,$deg. Using the flare energy estimation method from \citet{Shibayama_2013}, and the temperature of the flare $T_{\mathrm{flare}}=10000\pm{2000}\,$K \citep{Shibayama_2013, Kowalski_2015, Howard_2020}, we estimated bolometric energies for the modulated flare and the flare corrected for the effects of the rotational modulation. The bolometric energy of the flare before the correction was $(1.15\pm 0.35)\times 10^{35}\,$erg and after the correction was $(3.99\pm 1.22)\times 10^{35}\,$erg. The flare lasted about $27\,$h which makes it (so far) the longest flare ever detected in TESS data (see Figure \ref{fig:allflares} and \citet{Pietras_2022, Gunther_2020, Zilu_2021}).

\section{Discussion}\label{sec:conclusion}

\begin{figure*}[ht!]
    \centering
    \includegraphics[width=\textwidth]{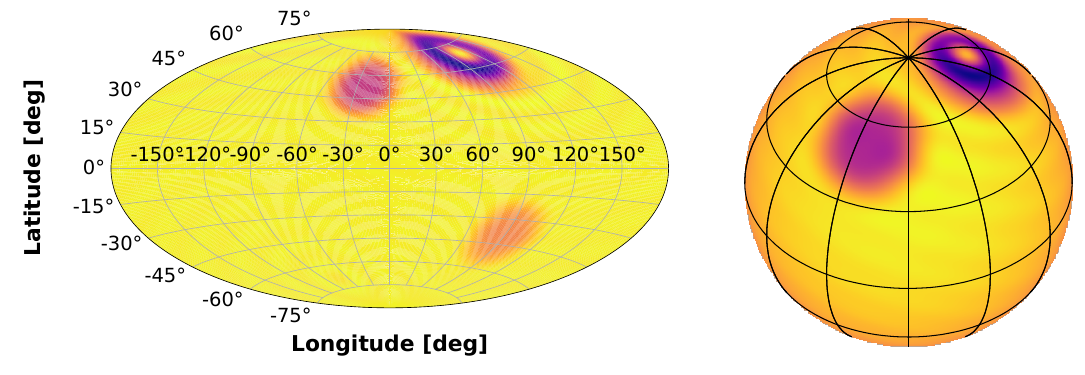}
    \caption{Both panels show the distribution of starspots on the star CD-36 3202 in sector 6 during the LDE flare. The bright region on the latitude approximately 70$^\circ$ and longitude approximately 100$^\circ$ is the flare from Figure \ref{fig:modulatedflare}. The left panel presents the star in the Aitoff projection and the right panel presents the model of the star in the orthographic projection.}
    \label{fig:flarespots}
\end{figure*}

We analyzed the starspot distribution on CD-36 3202, estimated the inclination angle of the star as equal to $70^\circ\pm8^\circ$, and modeled the light curve of this star to obtain the rotationally modulated LDE and estimate its parameters. It was possible to reconstruct a three-spot model with a mean spot temperature of approximately $4000\pm 765\,$K and average spottedness of about $11.61\%\pm0.13\,$\% in sector 6 of TESS observations. The LDE lasted approximately 27$\,$h, had a radius about $3.^{\!\!\circ}14$, had energy about $(3.99\pm 1.22)\times 10^{35}\,$erg, and occurred on $69^{+2}_{-1}\,$deg. For the first time, we were able to estimate the exact location of the flare comparing with the positions and sizes of the spots. The flaring region in the neighborhood of the spot is about $12^\circ$ from the center of the spot number 3 (see Figure \ref{fig:flarespots}. The white-light flare that occurs near the active region with the spots exhibits very solar-like behavior which is well documented on many images of the Sun from SDO/HMI \citep{Namekata_2017}. The dependency between the flares' occurrence and the starspots on K stars was only merely analyzed in the past. \citet{Roettenbacher_2018} showed that flares on K and M stars with amplitudes less than 5\% may be correlated with the groups of spots that are creating the minima in the light curve. Unfortunately, some authors show that for the M stars, this correlation may not occur \citep{Bicz_2022,Feinstein_2020,Ilin_2022}. Now, we can confirm that on K stars there is a chance for a flare to be in the starspot region and we suggest that this type of stars (partially convective flaring stars) should be the target of more observations (for example for the TESS mission) to analyze this case deeper. 

The flare's location estimated to $69^{+2}_{-1}\,$deg is similar to the results of other authors. \citet{Ilin_2021} analyzed four flares, modulated by the stellar rotation, on four late-type M stars. They showed that the flares on fast-rotating (rotational periods from 2.7$\,$h up to 8.4$\,$h) M stars tend to appear at significantly higher latitudes compared with the Sun. On the Sun flares occur in a belt around the equator up to the $30^\circ$. The flares found by \citet{Ilin_2021} were located on significantly higher latitudes above $55^\circ$. Our analyzed flare, which also is above the mentioned border, shows that this case may also be present not only in fully convective stars but also in partially convective stars. This may show similarities of the magnetic dynamo for young, rapidly rotating stars with different inner structures. The optical flux from the analyzed flare emitted towards hypothetical planets orbiting this star (or a similar star), assuming the spin-orbit alignment of the planets, decreases by $29^\circ\pm 12^\circ$. This value fits quite well the relation of relative optical flare energy emitted towards the planet in a star-planet system with a spin-orbit aligned for a typical superflare from \citet{Ilin_2021}. It is worth noticing that this result may impact only the white-light emission. The stellar coronae are optically thin so, the X-ray photons originating from flares may not be affected by the foreshortening effects similar to the effect that impacts the optical emission near the stellar limb. The X-ray flux received by the planets may therefore be much less attenuated.

\begin{figure*}[!b]
    \centering
    \includegraphics[width=\textwidth]{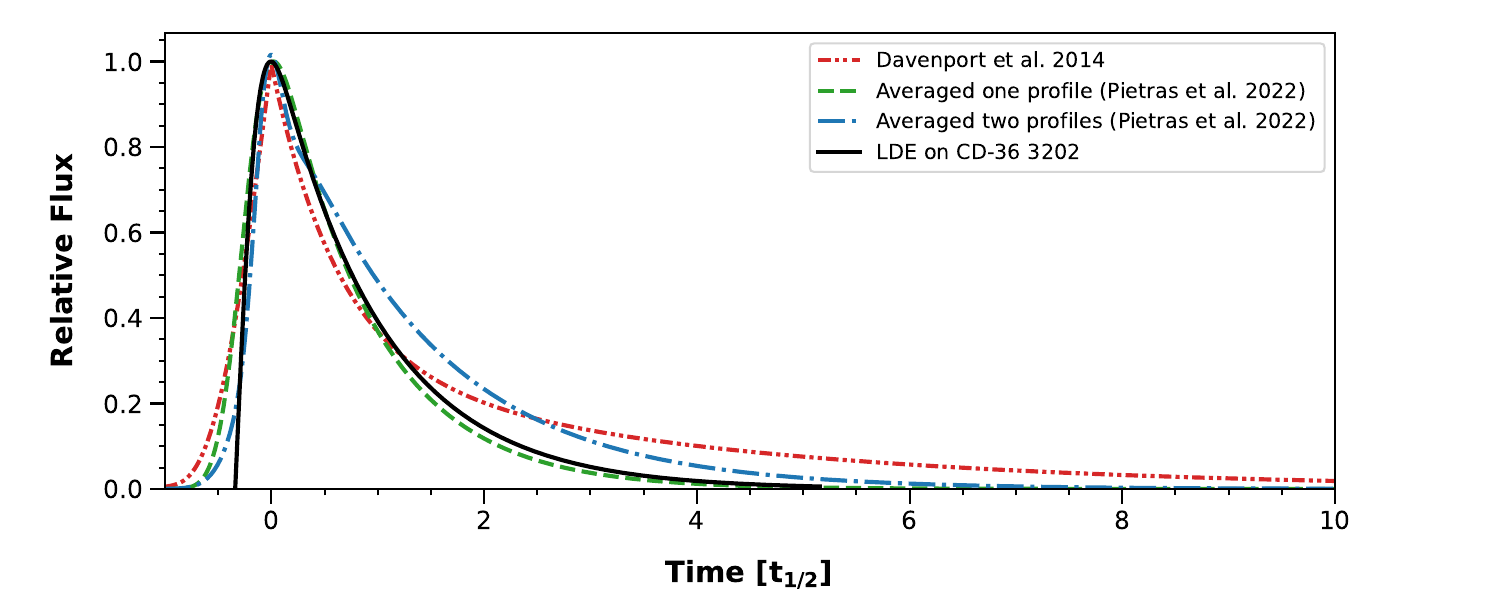}
    \caption{Comparision of averaged flare profiles from \citet{Davenport_2014} and \citet{Pietras_2022} with the profile of analyzed long-duration event. Red curve presents average flare profile presented by \citet{Davenport_2014}. The blue and green curves present average flare profiles presented by \citet{Pietras_2022}. They correspond to the one fitted profile (green curve) and the profile made as a combination of two profiles (blue curve). The black dashed curve corresponds to the analyzed modulated flare.}
    \label{fig:profiledav}
\end{figure*}

The gray data points visible in Figure \ref{fig:modulatedflare} (around 1487.1 days) that were excluded from the fitting process, indicate a decrease in the signal during the flare. This decrease is later repeated in the corresponding parts of the modulated flare (times about 1487.31 days and 1487.52 days). It can be seen that the moments of the beginning of the dimmings are separated in time by about 0.21$\,$day and are becoming shorter in time (107 minutes gray points, 30 minutes second dimming, 20 minutes third dimming). This can be caused by an optically thick absorption structure like a filament or a relatively cool loop that partially or completely covers the flaring region. 
The loops above the flaring region slowly change their visible structure due to the evolution of the temperature and density of the loops \citep{Petr_H, Song_2016, Ashwanden_book}. The areas covered by flare-loop arcades are large and comparable to the size of active regions, which is well documented on many images of the Sun from SDO/AIA. In the case of stellar flares, the loop arcades can be even much more extended. The other explanation may be the destabilization of the structure above the flaring region due to the flare and the eruption as a slow CME. Due to the CME's motion, the smaller part of the flaring region is covered.  
For a simple estimation of the plasma parameters of the covering structure we employed the cloud model \citep{Petr_H}. We hypothesized that the observed decrease (about 45\% of the emission of the flaring region) is caused by the absorbing structure above the flaring region. The decrease occurred near the peak of the flare emission in the TESS bandpass so we assumed that the temperature of the flaring region (for example flaring ribbons) during this period was 12,000 K. Additionally, we posited the temperature of the overlying cloud to be around 8,000 K (for example \citet{Jejcic_2018}) in order to achieve a reasonable temperature for the cooler structure. The flaring region has a radius about $3.^{\!\!\circ}14\,$deg which corresponds to $\sim\!\! 30,500\,$km size on the star. We approximate the cloud as an arcade of the post-flare loops that are semi-circular shape with a radius of $30,500\,$km. Using the radius of the flare loop and using the data about flare loops from \citet{Warmuth} we were able to estimate the loop cross-section radius as equal to approximately 2800$\,$km. Based on these parameters from a single loop, we were able to estimate that approximately 11 loops would be required to completely cover the region below them. We try to fit the absorption dip by varying the loop electron density using the covering structure, with an assumed temperature of $8000\,$K and a thickness of $5600\,$km. The resulting density we obtained is $7\times 10^{13}\,$cm$^{-3}$. The entire absorbing structure boasts a volume of roughly $3\times10^{28}\,$cm$^3$. A high electron density in the range of $n_e \geq  10^{12} - 10^{13}\,$cm$^{- 3}$ is both unquestionably necessary and anticipated, primarily due to the intense evaporative processes occurring during superflares (see discussion in \citet{Petr_H} ). While such high coronal densities have indeed been reported for some solar flares \citep{Hiei_1982,Jejcic_2018}. The complexity of this issue combined with the numerous degrees of freedom requires further research to thoroughly investigate the proposed hypotheses.

\begin{figure*}[!b]
    \centering
    \includegraphics[width=\textwidth]{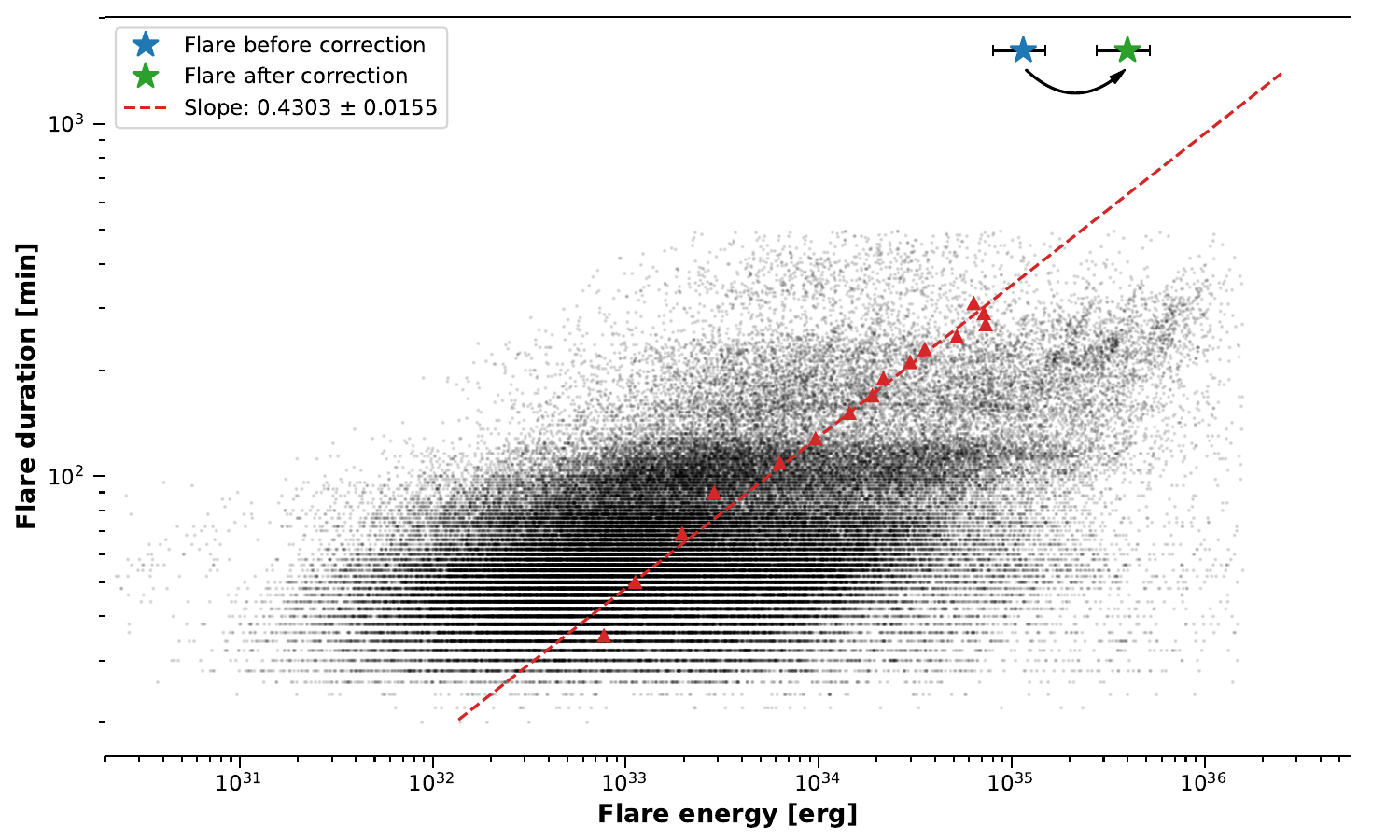}
    \caption{Observed relations between flare energy and duration time. All of the flares detected in the TESS data by \citet{Pietras_2022} are marked with black dots. The discrete flare duration bins result from the time resolution of the observations equal to 2 minutes. The red triangles show the location of the center of mass and the red dashed line presents the power-law fit. The blue star shows the energy of the analyzed flare before the correction for the rotational modulation and the green star after the correction. The black arrow presents the shift in the energy between modulated and demodulated flare.}
    \label{fig:allflares}
\end{figure*}

Before the modulated flare, at the TBJD about 1486.85 days was another flare. This flare's energy was $(5.31\pm 2.03)\times 10^{34}\,$erg. We tried to fit this event also as a part of the modulated flare. It was not possible because the time interval between the corresponding parts of the modulations (like local minima or local maxima) in the modulated flare should be separated in time by the rotational period of the star. Including this flare implemented a 53-minute time bias to the result making it impossible to do the fit correctly. This suggests that this flare was not part of the main event. The flare could be a precursor flare that led to the destabilization of great magnetic structure on the star and caused massive, long-lasting magnetic reconnection that led to an approximately 27-hour-long stellar flare. Another explanation is that this flare is just a random flare that was not connected in any way with the mentioned LDE.

We compared the profile of the flare with the mean profile of the flare presented by \citet{Davenport_2014} and with the mean profiles presented by \citet{Pietras_2022} (see Figure \ref{fig:profiledav}). To include both the impulsive rise and the slow exponent decay we used the time metric presented by \citet{Kowalski_2013}. We measured the light curve full-time width at half the maximum flux, denoted $t_{1/2}$. The flare profile obtained during our analysis fits very well with the averaged one-component profile estimated by \citet{Pietras_2022}. The flare on this star, if measured in $t_{1/2}$, has a faster rise time and slightly slower decay time than the averaged one profile (see Figure \ref{fig:profiledav}). Even though the duration time of this flare (approximately $27\,$h) is longer than any flare event observed on any star in TESS data \citep{Pietras_2022, Gunther_2020, Zilu_2021} it looks very similar to the averaged profile (average profile was made from profiles of about 94,000 flares). This may suggest that the mechanism of the flares - magnetic reconnection - is a mechanism that gives similar events but the energy is released in different timescales.

Also, we compared the bolometric energy of the analyzed flare and its duration with the corresponding parameters of over 140,000 flares detected on more than 25,000 stars \citep{Pietras_2022}. The energies of the flare, both before and after correction for rotational modulation effects, fall within the scatter range of all the flares (see Figure \ref{fig:allflares}). Additionally, the parameters of the corrected flare (the green star in Figure \ref{fig:allflares}) position it noticeably closer to the exponential fit observed for all flares (the red line in Figure \ref{fig:allflares}). Similarly, the energy of the flare and the starspot area where the flare occurred ($3.93\pm0.3$ percent of the stellar surface) fits very well to the relation flare energy$\,$-$\,$spot group area for superflares on solar-type stars presented by \citet{Maehara_2015}. These findings suggest that the energy release mechanism for this particular flare may align with the typical process seen in white-light flares - the heating caused by non-thermal electrons accelerated through magnetic reconnection \citep{Brown_1971, Emsile_1978} but the energy is released in much higher timescale than in the mostly observed stellar and solar flares \citep{Pietras_2022, Gunther_2020}.

\section{Acknowledgements}
This work was partially supported by the program "Excellence Initiative - Research University" for years 2020-2026 for the University of Wrocław, project no. BPIDUB.4610.96.2021.KG. The computations were performed using resources provided by Wrocław Networking and Supercomputing Centre\footnote{https://wcss.pl}, computational grant number 569. We extend our gratitude to Professor Petr Heinzel for the invaluable discussions regarding the role of flare loops in the white light emission of stellar flares and for his help with cloud-model computations. The authors are grateful to the anonymous referee for constructive comments and suggestions, which have proved to be very helpful in improving the manuscript. This paper includes data collected by the TESS mission. Funding for the TESS mission is provided by NASA's Science Mission Directorate.

\bibliography{biblio}{}

\begin{thebibliography}{47}
\expandafter\ifx\csname natexlab\endcsname\relax\def\natexlab#1{#1}\fi

\bibitem[{{Aschwanden}(2005)}]{Ashwanden_book}
{Aschwanden}, M.~J. 2005, {Physics of the Solar Corona. An Introduction with
  Problems and Solutions (2nd edition)}

\bibitem[{Bell {et~al.}(2015)Bell, Mamajek, \& Naylor}]{bell_2014}
Bell, C. P.~M., Mamajek, E.~E., \& Naylor, T. 2015, Monthly Notices of the
  Royal Astronomical Society, 454, 593

\bibitem[{Bicz {et~al.}(2022)Bicz, Falewicz, Pietras, Siarkowski, \&
  Preś}]{Bicz_2022}
Bicz, K., Falewicz, R., Pietras, M., Siarkowski, M., \& Preś, P. 2022, The
  Astrophysical Journal, 935, 102

\bibitem[{{Biermann}(1941)}]{Biermann_1941}
{Biermann}, L. 1941, Vierteljahresschrift der Astronomischen Gesellschaft, 76,
  194

\bibitem[{Bogner {et~al.}(2022)Bogner, Stelzer, \& Raetz}]{Bogner_2022}
Bogner, M., Stelzer, B., \& Raetz, S. 2022, Astronomische Nachrichten, 343,
  e210079

\bibitem[{{Bray} \& {Loughhead}(1964)}]{Bray_and_Loughhead_1964}
{Bray}, R.~J. \& {Loughhead}, R.~E. 1964, {Sunspots}

\bibitem[{{Brown}(1971)}]{Brown_1971}
{Brown}, J.~C. 1971, \solphys, 18, 489

\bibitem[{Chitre(1963)}]{Chitre_1963}
Chitre, S.~M. 1963, Monthly Notices of the Royal Astronomical Society, 126, 431

\bibitem[{{Claret, A.}(2017)}]{Claret_2017}
{Claret, A.} 2017, A\&A, 600, A30

\bibitem[{Davenport {et~al.}(2019)Davenport, Covey, Clarke, Boeck, Cornet, \&
  Hawley}]{Davenport_2019}
Davenport, J. R.~A., Covey, K.~R., Clarke, R.~W., {et~al.} 2019, The
  Astrophysical Journal, 871, 241

\bibitem[{Davenport {et~al.}(2014)Davenport, Hawley, Hebb, Wisniewski,
  Kowalski, Johnson, Malatesta, Peraza, Keil, Silverberg, Jansen, Scheffler,
  Berdis, Larsen, \& Hilton}]{Davenport_2014}
Davenport, J. R.~A., Hawley, S.~L., Hebb, L., {et~al.} 2014, The Astrophysical
  Journal, 797, 122

\bibitem[{{Deinzer}(1965)}]{Deinzer_1965}
{Deinzer}, W. 1965, \apj, 141, 548

\bibitem[{{Dicke}(1970)}]{Dicke_1970}
{Dicke}, R.~H. 1970, \apj, 159, 25

\bibitem[{{Emslie}(1978)}]{Emsile_1978}
{Emslie}, A.~G. 1978, \apj, 224, 241

\bibitem[{Feinstein {et~al.}(2020)Feinstein, Montet, Ansdell, Nord, Bean,
  Günther, Gully-Santiago, \& Schlieder}]{Feinstein_2020}
Feinstein, A.~D., Montet, B.~T., Ansdell, M., {et~al.} 2020, The Astronomical
  Journal, 160, 219

\bibitem[{{Feoktistov} \& {Janaqi}(2004)}]{Feoktistov_2004}
{Feoktistov}, V. \& {Janaqi}, S. 2004, {Proceedings of the 18th International
  Parallel and Distributed Processing Symposium (IPDPS’04)}, 165

\bibitem[{Fletcher(1987)}]{Flet87}
Fletcher, R. 1987, Practical Methods of Optimization, 2nd edn. (New York, NY,
  USA: John Wiley \& Sons)

\bibitem[{Flores {et~al.}(2022)Flores, Connelley, Reipurth, \&
  Duchêne}]{Flores_2022}
Flores, C., Connelley, M.~S., Reipurth, B., \& Duchêne, G. 2022, The
  Astrophysical Journal, 925, 21

\bibitem[{Foreman-Mackey {et~al.}(2020)Foreman-Mackey, Luger, Czekala, Agol,
  Price-Whelan, Brandt, Barclay, \& Bouma}]{exoplanet:exoplanet}
Foreman-Mackey, D., Luger, R., Czekala, I., {et~al.} 2020,
  exoplanet-dev/exoplanet v0.4.0

\bibitem[{{Gryciuk} {et~al.}(2017){Gryciuk}, {Siarkowski}, {Sylwester},
  {Gburek}, {Podgorski}, {Kepa}, {Sylwester}, \& {Mrozek}}]{Gryciuk_2017}
{Gryciuk}, M., {Siarkowski}, M., {Sylwester}, J., {et~al.} 2017, \solphys, 292,
  77

\bibitem[{{G{\"u}nther} {et~al.}(2020){G{\"u}nther}, {Zhan}, {Seager},
  {Rimmer}, {Ranjan}, {Stassun}, {Oelkers}, {Daylan}, {Newton}, {Kristiansen},
  {Olah}, {Gillen}, {Rappaport}, {Ricker}, {Vanderspek}, {Latham}, {Winn},
  {Jenkins}, {Glidden}, {Fausnaugh}, {Levine}, {Dittmann}, {Quinn},
  {Krishnamurthy}, \& {Ting}}]{Gunther_2020}
{G{\"u}nther}, M.~N., {Zhan}, Z., {Seager}, S., {et~al.} 2020, \aj, 159, 60

\bibitem[{{Heinzel} \& {Shibata}(2018)}]{Petr_H}
{Heinzel}, P. \& {Shibata}, K. 2018, \apj, 859, 143

\bibitem[{{Hiei}(1982)}]{Hiei_1982}
{Hiei}, E. 1982, \solphys, 80, 113

\bibitem[{Howard {et~al.}(2020)Howard, Corbett, Law, Ratzloff, Galliher,
  Glazier, Gonzalez, Soto, Fors, del Ser, \& Haislip}]{Howard_2020}
Howard, W.~S., Corbett, H., Law, N.~M., {et~al.} 2020, The Astrophysical
  Journal, 902, 115

\bibitem[{{Howard} {et~al.}(2019){Howard}, {Corbett}, {Law}, {Ratzloff},
  {Glazier}, {Fors}, {del Ser}, \& {Haislip}}]{Howard_2019}
{Howard}, W.~S., {Corbett}, H., {Law}, N.~M., {et~al.} 2019, \apj, 881, 9

\bibitem[{{Hoyle}(1949)}]{Hoyle_1949}
{Hoyle}, F. 1949, {Some recent researches in solar physics.}

\bibitem[{Ilin \& Poppenhaeger(2022)}]{Ilin_2022}
Ilin, E. \& Poppenhaeger, K. 2022, Monthly Notices of the Royal Astronomical
  Society, 513, 4579

\bibitem[{Ilin {et~al.}(2021)Ilin, Poppenhaeger, Schmidt, Järvinen, Newton,
  Alvarado-Gómez, Pineda, Davenport, Oshagh, \& Ilyin}]{Ilin_2021}
Ilin, E., Poppenhaeger, K., Schmidt, S.~J., {et~al.} 2021, Monthly Notices of
  the Royal Astronomical Society, 507, 1723

\bibitem[{{Jej{\v{c}}i{\v{c}}} {et~al.}(2018){Jej{\v{c}}i{\v{c}}}, {Kleint}, \&
  {Heinzel}}]{Jejcic_2018}
{Jej{\v{c}}i{\v{c}}}, S., {Kleint}, L., \& {Heinzel}, P. 2018, \apj, 867, 134

\bibitem[{{Kov{\'a}ri} {et~al.}(2007){Kov{\'a}ri}, {Vilardell}, {Ribas},
  {Vida}, {van Driel-Gesztelyi}, {Jordi}, \& {Ol{\'a}h}}]{Kovari_2007}
{Kov{\'a}ri}, Z., {Vilardell}, F., {Ribas}, I., {et~al.} 2007, Astronomische
  Nachrichten, 328, 904

\bibitem[{{Kowalski} {et~al.}(2015){Kowalski}, {Hawley}, {Carlsson}, {Allred},
  {Uitenbroek}, {Osten}, \& {Holman}}]{Kowalski_2015}
{Kowalski}, A.~F., {Hawley}, S.~L., {Carlsson}, M., {et~al.} 2015, \solphys,
  290, 3487

\bibitem[{{Kowalski} {et~al.}(2013){Kowalski}, {Hawley}, {Wisniewski}, {Osten},
  {Hilton}, {Holtzman}, {Schmidt}, \& {Davenport}}]{Kowalski_2013}
{Kowalski}, A.~F., {Hawley}, S.~L., {Wisniewski}, J.~P., {et~al.} 2013, \apjs,
  207, 15

\bibitem[{Luger {et~al.}(2019)Luger, Agol, Foreman-Mackey, Fleming,
  Lustig-Yaeger, \& Deitrick}]{Luger_2019}
Luger, R., Agol, E., Foreman-Mackey, D., {et~al.} 2019, The Astronomical
  Journal, 157, 64

\bibitem[{{Maehara} {et~al.}(2015){Maehara}, {Shibayama}, {Notsu}, {Notsu},
  {Honda}, {Nogami}, \& {Shibata}}]{Maehara_2015}
{Maehara}, H., {Shibayama}, T., {Notsu}, Y., {et~al.} 2015, Earth, Planets and
  Space, 67, 59

\bibitem[{{Mathur, S.} {et~al.}(2014){Mathur, S.}, {Garc\'{\i}a, R. A.},
  {Ballot, J.}, {Ceillier, T.}, {Salabert, D.}, {Metcalfe, T. S.}, {R\'egulo,
  C.}, {Jim\'enez, A.}, \& {Bloemen, S.}}]{Mathur_2014}
{Mathur, S.}, {Garc\'{\i}a, R. A.}, {Ballot, J.}, {et~al.} 2014, A\&A, 562,
  A124

\bibitem[{{Namekata} {et~al.}(2017){Namekata}, {Sakaue}, {Watanabe}, {Asai},
  {Maehara}, {Notsu}, {Notsu}, {Honda}, {Ishii}, {Ikuta}, {Nogami}, \&
  {Shibata}}]{Namekata_2017}
{Namekata}, K., {Sakaue}, T., {Watanabe}, K., {et~al.} 2017, \apj, 851, 91

\bibitem[{Pietras {et~al.}(2022)Pietras, Falewicz, Siarkowski, Bicz, \&
  Preś}]{Pietras_2022}
Pietras, M., Falewicz, R., Siarkowski, M., Bicz, K., \& Preś, P. 2022, The
  Astrophysical Journal, 935, 143

\bibitem[{Ricker {et~al.}(2014)Ricker, Winn, Vanderspek, Latham, Bakos, Bean,
  Berta-Thompson, Brown, Buchhave, Butler, Butler, Chaplin, Charbonneau,
  Christensen-Dalsgaard, Clampin, Deming, Doty, Lee, Dressing, Dunham, Endl,
  Fressin, Ge, Henning, Holman, Howard, Ida, Jenkins, Jernigan, Johnson,
  Kaltenegger, Kawai, Kjeldsen, Laughlin, Levine, Lin, Lissauer, MacQueen,
  Marcy, McCullough, Morton, Narita, Paegert, Palle, Pepe, Pepper, Quirrenbach,
  Rinehart, Sasselov, Sato, Seager, Sozzetti, Stassun, Sullivan, Szentgyorgyi,
  Torres, Udry, \& Villasenor}]{Ricker_2014}
Ricker, G.~R., Winn, J.~N., Vanderspek, R., {et~al.} 2014, Space Telescopes and
  Instrumentation 2014: Optical, Infrared, and Millimeter Wave, 9143, 914320

\bibitem[{Roettenbacher \& Vida(2018)}]{Roettenbacher_2018}
Roettenbacher, R.~M. \& Vida, K. 2018, The Astrophysical Journal, 868, 3

\bibitem[{Salvatier {et~al.}(2016)Salvatier, Wiecki, \&
  Fonnesbeck}]{Salvatier_2016}
Salvatier, J., Wiecki, T.~V., \& Fonnesbeck, C. 2016, {PeerJ} Computer Science,
  2, e55

\bibitem[{Shibayama {et~al.}(2013)Shibayama, Maehara, Notsu, Notsu, Nagao,
  Honda, Ishii, Nogami, \& Shibata}]{Shibayama_2013}
Shibayama, T., Maehara, H., Notsu, S., {et~al.} 2013, The Astrophysical Journal
  Supplement Series, 209, 5

\bibitem[{Song {et~al.}(2016)Song, Wang, Feng, \& Zhang}]{Song_2016}
Song, Q., Wang, J.-S., Feng, X., \& Zhang, X. 2016, The Astrophysical Journal,
  821, 83

\bibitem[{{Theano Development Team}(2016)}]{exoplanet:theano}
{Theano Development Team}. 2016, arXiv e-prints, abs/1605.02688

\bibitem[{{Torres, C. A. O.} {et~al.}(2006){Torres, C. A. O.}, {Quast, G. R.},
  {da Silva, L.}, {de la Reza, R.}, {Melo, C. H. F.}, \& {Sterzik,
  M.}}]{Torres_2006}
{Torres, C. A. O.}, {Quast, G. R.}, {da Silva, L.}, {et~al.} 2006, A\&A, 460,
  695

\bibitem[{{Warmuth} \& {Mann}(2013)}]{Warmuth}
{Warmuth}, A. \& {Mann}, G. 2013, \aap, 552, A87

\bibitem[{Yang {et~al.}(2017)Yang, Liu, Gao, Fang, Guo, Zhang, Hou, Wang, \&
  Cao}]{Yang_2017}
Yang, H., Liu, J., Gao, Q., {et~al.} 2017, The Astrophysical Journal, 849, 36

\bibitem[{{Yang, Zilu} {et~al.}(2023){Yang, Zilu}, {Zhang, Liyun}, {Meng,
  Gang}, {Han, Xianming L.}, {Misra, Prabhakar}, {Yang, Jiawei}, \& {Pi,
  Qingfeng}}]{Zilu_2021}
{Yang, Zilu}, {Zhang, Liyun}, {Meng, Gang}, {et~al.} 2023, A\&A, 669, A15

\end{thebibliography}
\bibliographystyle{aa}

\begin{appendix}
\section{CD-36 3202 Inclination Angle Estimation}\label{appendix:inc}
\begin{figure*}[!b]
    \centering
    \includegraphics[width=\textwidth]{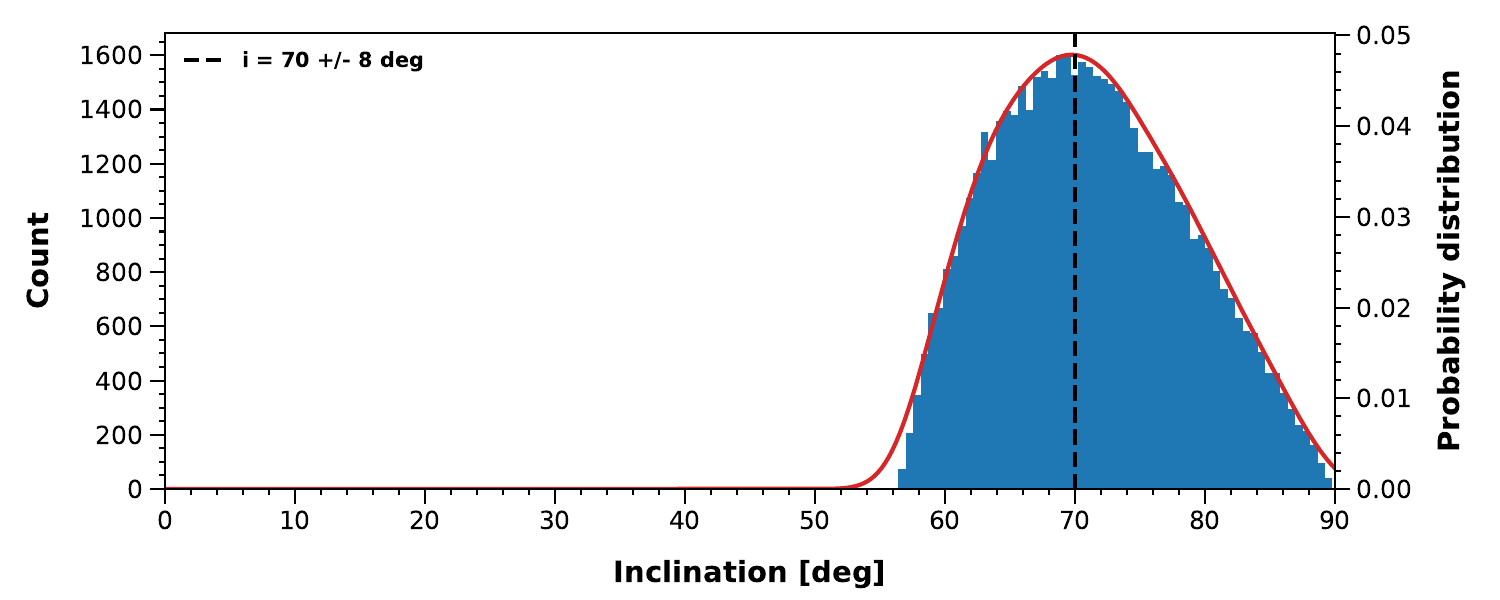}
    \caption{Posterior distribution of stellar inclination angles (blue histogram), the smoothed probability density posterior (red curve), the maximum of the posterior (black dotted line).}
    \label{fig:inclination}
\end{figure*}

We utilized the stellar rotation period, $P_*$, and radius, $R_*$, in conjunction with the projection of rotational velocity $v\sin (i)$, to deduce the stellar inclination angle, $i$. To address the statistical correlation between equatorial velocity and $v\sin (i)$, we employed our newly developed tool, \texttt{findinc\_mc}\footnote{https://github.com/KBicz/findincmc}. This software employs Equation \ref{eq:vsini} and Monte-Carlo techniques to generate a probability density posterior for the inclination angle based on the provided measurements.
\begin{equation}\label{eq:vsini}
 \hfil\hfil i = \sin^{-1}\left(\frac{vsin(i)\cdot P_*}{2\pi R_*}\right).
\end{equation}
We imposed the condition $0 < \sin (i) < 1$ while assuming Gaussian priors for $v\sin (i)$, $P_*$, and $R_*$. The inclination posteriors exhibit non-Gaussian behavior across various stars, although they can be approximated by a double Gaussian fit. The inclination value corresponds to the angle that maximizes the distribution. For uncertainty estimation, we calculate the standard deviation of the posterior distribution. Using the \texttt{findinc\_mc} software we evaluated the inclination of the star as equal to $70^\circ\pm 8^\circ$ (see Figure \ref{fig:inclination}). During the calculation process we used the estimated rotational period equal $P_* = 0.23536 \pm 0.00046\,$day and we assumed the projection of rotation velocity $vsin(i)=170\pm 17\,$km/s \citep{Torres_2006}. The radius of the star $R_*$ was taken from the MAST catalog.

\end{appendix}

\end{document}